\documentstyle[12pt,aasms4]{article} 
\begin{document} 
\noindent {\it Dissertation Summary} 
\begin{center} 
\title{\large \bf Mass Transfer Processes in Classical Nova Systems
 Around Their Outburst Events and Beyond} 
\end{center} 
\author{Alon Retter} 
\affil{Physics Dept., Keele University, Staffordshire, ST5 5BG, UK}
\begingroup \parindent=1cm 
\begin{center} Electronic mail: ar@astro.keele.ac.uk;
Thesis work conducted at: Dept. of Physics and Astron., Tel Aviv University,
Tel Aviv, 69978, Israel;
Ph.D. Thesis directed by: Prof. Elia Leibowitz;
 ~Ph.D. Degree awarded: 1999
\end{center} \endgroup 
A continuous photometric study of 12 novae was carried out for more than 
365 nights during the past four years at the Wise Observatory in a 
motivation to extend our knowledge on these systems. Table 1 presents 
a summary of the observations and the main results.

\begin{table}
\caption {The Observations and Main Results}
\begin{tabular}{@{}lcccc@{}}
\\
\hline

Nova      &Number of& Period / Results     & Classification& Notes \\
          & Nights  &                      &               & \\
\hline

Aql 1993  & 21      & photometric-?        &               & \\

Aql 1995  & 31      & orbital+spin+beat    & intermediate polar &MNRAS paper \\

V1370 Aql & 16      & ---------            &               & short runs \\

Cas 1993  & 55 (+6) & orbital              &               & co. P. Szkody\\
          &         &                      &               & MNRAS preprint\\

Cas 1995  & 26      & irregular variations &               & co. O. Shemmer \\

Cyg 1992  & 80 (+34)& orbital + superhump  & permanent superhump & 2 MNRAS papers + \\
          &         &                      & system        & international campaign\\   
          &         &                      &               & (1997 July)\\

DM Gem    & 29      & irregular variations &               & co. Y. Lipkin\\

DN Gem    & 39      & orbital              & irradiation effect & MNRAS paper \\

Oph 1994  & 10      & photometric-?        &               & \\

Sgr 2 1994& 2       & --------             &               & \\

Sgr 1998  & 28      & photometric          &               & co. Y. Lipkin  \\

RW UMi    & 30      & irregular variations &               & \\

\hline

\end{tabular}

\end{table}

It was believed that the accretion disc is destroyed by the nova outburst,
and that it takes only a few decades for the disc to reform. The main aim 
of the project was, therefore, to try to find observational evidence for
the presence of an accretion disc within the binary nova system shortly
after its outburst event. Positive results were found in one case 
(permanent superhumps in Nova V1974 Cygni 1992 - Retter A., Leibowitz E.M., 
Ofek E.O., 1997, MNRAS, 286, 745), and most probably in another case 
(the classification of Nova V1425 Aql 1995 as an intermediate polar system - 
Retter A., Leibowitz E.M., Kovo-Kariti O., 1998, MNRAS, 293, 145). In 
addition, we used the Eccentric Disc model in order to derive the binary 
parameters of Nova V1974 Cygni 1992 (Retter et al. 1997).

Further use of the theory and simple assumptions lead to a prediction
for the future of superhumps in classical nova systems, and in particular 
for the permanent superhumps in Nova V1974 Cygni 1992 (Retter A., Leibowitz 
E.M., 1998, MNRAS, 296, L37). They will either continue to prevail in 
its light curve, while the brightness of the nova stays well above its 
pre-outburst level, or the system will continue to decay and eventually 
evolve into a regular SU UMa system, with superhumps appearing only during 
superoutbursts.

The detection of permanent superhumps in Nova Cygni 1992 and the observed 
evolution of two other old classical novae - V603 Aquilae 1918 and CP Puppis 
1942 - lead us to suggest that non-magnetic short-period nova systems are the 
progenitors of permanent superhump systems, which later in their life-time 
might evolve into SU UMa systems (Retter \& Leibowitz, 1998).

We also found a photometric period in the light curve of Nova DN Gem 1912,
and proposed that it is the orbital period of the binary system (Retter A., 
Leibowitz E.M., Naylor T., 1999, MNRAS, accepted). This interpretation 
makes DN Gem the fourth nova inside the period gap distribution of 
cataclysmic variables, and it bolsters the idea that there is no gap for 
classical novae. However, the small number of known nova periods is still 
too poor to establish this idea statistically. We further argued that the 
modulation is driven by an irradiation effect.

We also detected periodic variations in Nova V705 Cas 1993 (Retter A., 
Leibowitz E.M., 1995, IAUC, 6234; Leibowitz E.M. et al., in preparation), 
and in Nova Sgr 1998 (Lipkin Y., Retter A., Leibowitz E.M., 1998, IAUC, 
6963). The observations of Nova Aql 1993 and Nova Oph 
1994 revealed a possible photometric period in each. The reduction and 
analysis of the data of both novae are under way. Irregular variations 
were found in the light curves of Nova Cas 1995, DM Gem and RW UMi. 
In 1997, I lead an international photometric campaign on Nova V1974 Cyg 
1992. These data are being analyzed.





\end{document}